# Supersymmetric Yang-Mills Theories in $D \geq 12$ [1]

Hitoshi NISHINO

*Department of Physics*
*University of Maryland*
*College Park, MD 20742-4111, USA*
E-Mail: nishino@umdhep.umd.edu

## Abstract

We present supersymmetric Yang-Mills theories in arbitrary even dimensions with the signature $(9+m, 1+m)$ where $m = 0, 1, 2, \cdots$ beyond ten-dimensions up to infinity. This formulation utilizes null-vectors and is a generalization of our previous work in 10+2 dimensions to arbitrary even dimensions with the above signature. We have overcome the previously-observed obstruction beyond 11+3 dimensions, by the aid of projection operators. Both component and superspace formulations are presented. This also suggests the possibility of consistent supergravity theories in any even dimensions beyond 10+1 dimensions.

---

[1]This work is supported in part by NSF grant # PHY-93-41926.

# 1. Introduction

Recently we have constructed an $N=1$ supersymmetric Yang-Mills (SYM) theory [1], an $N=1$ supergravity theory [2], and an $N=2$ supergravity theory [3] in twelve-dimensions with the signature $(10,2)$,[2] motivated by the development of F-theory in 12D [4][5][6], S-theory [7], or theories with two times [8]. There is some indication that the perturbative and non-perturbative states of M-theory may admit a unification within the framework of a superalgebra in $D=(10,2)$ [9] or $D=(11,3)$ [10]. In fact, an explicit SYM theory in $D=(11,3)$ has been constructed in [11]. However, it was also reported [11] that there is an obstruction to construct SYM beyond $D=(11,3)$, and therefore it was concluded that $D=(11,3)$ may be the *maximal* dimensions for SYM. The main obstruction was that the supersymmetric variation of the gaugino field equation seemed to produce unwanted $\gamma$-matrix structures which can not be absorbed as the gauge field equation [11]. Other than this problem, there seemed to be no additional problem, such as the on-shell closure of algebra on all the field, or the consistency of extra constraints with supersymmetry.

However, such an isolated obstruction [11] seems unusual to our past experience with supersymmetry, because once the closure of gauge algebra works on all the fields by help of fermionic field equations, there usually arises no inconsistency at the field equation level. We look into this obstruction with SYM theory in dimensions higher than $D=(11,3)$, and we overcome the problem about the unwanted term by using projection operators. We find that those undesirable terms in the variation of the gaugino field equations [11] vanish due to the property of the projection operators. As by-products, we generalize this result to arbitrary higher even dimensions, establishing both component and superspace formulations of SYM theories in arbitrary even dimensions with the signature $D=(9+m, 1+m)$ with $m=0,1,2,\cdots$ beyond $D=(9,1)$ up to infinity.

We first categorize the symmetries of gamma-matrices [12] depending on the space-time dimensions, all with eight dimensional difference between spacial and time coordinates, into four classes: $D=(9+4n, 1+4n)$, $D=(10+4n, 2+4n)$, $D=(11+4n, 3+4n)$, $D=(12+4n, 4+4n)$ for an arbitrary non-negative integer $n$, and analyze each case in a separate section. We take the advantage of the fact that the symmetry property of the gamma-matrices conveniently repeats itself every 8 dimensions in space-time [12], while the chirality (dottedness) of the spinors alternates every two dimensions. For instance, the dottedness of gamma-matrices in $D=(9+4n, 1+4n)$ is similar to that in $D=(11+4n, 3+4n)$, while $D=(10+4n, 2+4n)$ similar to $D=(12+4n, 4+4n)$. We present the component formulations in these dimensions in separate sections, with corresponding superspace formulations.

---

[2]Here the number 10 denotes the number of space-dimensions, while 2 is for that of time directions. We symbolize this space-time by $D=(10,2)$ from now on.

## 2. SYM in $D = (9+4n, 1+4n)$

As the first series of our SYM theories, we investigate dimensions with $9+4n$ space and $1+4n$ time coordinates, where $n = 0, 1, 2, \cdots$. Starting with the component formulation result, we come to the question of obstruction [11], and subsequently we will give the corresponding superspace formulation.

The structure of gamma-matrices in $D = (9+4n, 1+4n)$ [12] is parallel to $D = (9,1)$, namely, $\gamma^{[4k+1]}$ and $\gamma^{[4k+2]}$ are symmetric, while $\gamma^{[4k]}$ and $\gamma^{[4k+3]}$ are antisymmetric:

$$\begin{aligned} \text{Symmetric:} \quad & \gamma^\mu, \quad \gamma^{\mu\nu}, \quad \cdots, \quad \gamma^{[4k+1]}, \quad \gamma^{[4k+2]}, \quad \cdots, \quad \gamma^{[10+8n]}, \\ \text{Antisymmetric:} \quad & C, \quad \gamma^{[3]}, \quad \cdots, \quad \gamma^{[4k+3]}, \quad \gamma^{[4k+4]}, \quad \cdots, \quad \gamma^{[8+8n]}. \end{aligned} \quad (2.1)$$

In this paper, we use the collective indices $e.g.$, $[k]$ for the totally antisymmetric indices $\mu_1 \cdots \mu_k$, saving considerable space: $U^{[k]} V_{[k]} \equiv U^{\mu_1 \cdots \mu_k} V_{\mu_1 \cdots \mu_k}$. Our metric is

$$\left(\eta_{\mu\nu}\right) = \text{diag.} \begin{pmatrix} 0 & 1 & \cdots & 9 & 11 & 12 & 13 & 14 & 15 & 16 & \cdots & 9+8n & 10+8n \\ - & + & \cdots & + & + & - & + & - & + & - & \cdots & + & - \end{pmatrix}. \quad (2.2)$$

Our field content is the same as the usual SYM theory in $D = (9,1)$, namely the real gauge field $A_\mu{}^I$ with the adjoint representation index $I$, and the Majorana-Weyl gaugino $\lambda^I$ also in the adjoint representation. Our transformation rule is

$$\begin{aligned} \delta_Q A_\mu{}^I &= (\overline{\epsilon} \gamma_\mu \lambda^I), \\ \delta_Q \lambda^I &= \tfrac{1}{4} \gamma^{\mu\nu[4n]} \epsilon F_{\mu\nu}{}^I v_{[4n]}. \end{aligned} \quad (2.3)$$

Here $v_{[4n]}$ is a totally antisymmetric product of all the null-vectors:

$$v_{\mu_1 \cdots \mu_{4n}} \equiv n^{(1)}_{[\mu_1} n^{(2)}_{\mu_2} \cdots n^{(4n)}_{\mu_{4n}]}, \quad (2.4)$$

where our null-vectors are defined for $i = 1, 2, \cdots, 4n$ by

$$\left(n^{(i)}_\mu\right) \equiv \begin{pmatrix} 0 & 1 & \cdots & 9 & 11 & 12 & \cdots & 9+2i & 10+2i & \cdots & 9+8n & 10+8n \\ 0, & 0, & \cdots, & 0, & 0, & 0, & \cdots, & +\tfrac{1}{\sqrt{2}}, & +\tfrac{1}{\sqrt{2}}, & \cdots, & 0, & 0 \end{pmatrix}. \quad (2.5)$$

The superscript $(i)$ corresponds to the $i$-th extra pairs of coordinates. It is useful to define another set of null-vectors by

$$\left(m^{(i)}_\mu\right) \equiv \begin{pmatrix} 0 & 1 & \cdots & 9 & 11 & 12 & \cdots & 9+2i & 10+2i & \cdots & 9+8n & 10+8n \\ 0, & 0, & \cdots, & 0, & 0, & 0, & \cdots, & +\tfrac{1}{\sqrt{2}}, & -\tfrac{1}{\sqrt{2}}, & \cdots, & 0, & 0 \end{pmatrix}. \quad (2.6)$$

It is also convenient to use the $\pm$-coordinates defined by

$$V^{(i)}_\pm \equiv \tfrac{1}{\sqrt{2}} \left(V_{9+2i} \pm V_{10+2i}\right) \quad (2.7)$$

for the extra coordinates, so that $n_+^{(i)} \equiv m_-^{(i)} \equiv +1$, *etc*. We define the projection operators

$$P_\uparrow^{(i)} \equiv \tfrac{1}{2} \slashed{n}^{(i)} \slashed{m}^{(i)} \ , \qquad P_\downarrow^{(i)} \equiv \tfrac{1}{2} \slashed{m}^{(i)} \slashed{n}^{(i)} \ , \tag{2.8}$$

where $\slashed{n}^{(i)} \equiv \gamma^\mu n_\mu^{(i)}$, $\slashed{m}^{(i)} \equiv \gamma^\mu m_\mu^{(i)}$, as a generalization of similar operators in [2]. In (2.8), we do not take any summation over $(i)$, as is self-explanatory. These projection operators satisfy the usual ortho-normality relations

$$P_\uparrow^{(i)} + P_\downarrow^{(i)} \equiv I \ , \quad P_\uparrow^{(i)} P_\downarrow^{(i)} = P_\downarrow^{(i)} P_\uparrow^{(i)} = 0 \ , \quad (P_\uparrow^{(i)})^2 = P_\uparrow^{(i)} \ , \quad (P_\downarrow^{(i)})^2 = P_\downarrow^{(i)} \ ,$$
$$[\, P_\uparrow^{(i)}, P_\uparrow^{(j)} \,] = [\, P_\downarrow^{(i)}, P_\downarrow^{(j)} \,] = [\, P_\uparrow^{(i)}, P_\downarrow^{(j)} \,] = 0 \ . \tag{2.9}$$

As in $D = (10,2)$ [1], the gauge field undergoes the extra transformation

$$\delta_E A_\mu{}^I = v_{\mu [\, 4n-1 \,]} \Omega^{[\, 4n-1 \,] I} \ , \tag{2.10}$$

with the parameter $\Omega^{[\, 4n-1 \,] I}$, and all of our fields obey the constraints

$$v_{[\, 4n-1 \,]}{}^\mu D_\mu \lambda^I = 0 \ , \tag{2.11}$$

$$v_{[\, 4n-1 \,]}{}^\mu \gamma_\mu \lambda^I = 0 \ , \tag{2.12}$$

$$v_{[\, 4n-1 \,]}{}^\mu F_{\mu\nu}{}^I = 0 \ . \tag{2.13}$$

We can also confirm the consistency of these constraints under supersymmetry (2.3).

Our field equations are

$$\gamma^\mu D_\mu \lambda^I = 0 \ , \tag{2.14}$$

$$D^\mu F_{\mu [\, \nu_1}{}^I v_{\nu_2 \cdots \nu_{4n+1}]} = \frac{1}{2^{4n}(4n+1)!} f^{IJK} (\overline{\lambda}^J \gamma_{\nu_1 \cdots \nu_{4n+1}} \lambda^K) \ . \tag{2.15}$$

Eq. (2.15) contains the most authentic SYM theory in $D = (9,1)$ for $n = 0$ as the simplest case.

The on-shell closure of our supersymmetry (2.3) is confirmed as in [1]:

$$[\, \delta_Q(\epsilon_1), \delta_Q(\epsilon_2) \,] = \delta_P(\xi) + \delta_G(\Lambda) + \delta_E(\Omega) \ , \tag{2.16}$$

with the translation $\delta_P$, gauge transformation $\delta_G$, and the extra transformation $\delta_E$ with the parameters

$$\xi^\mu \equiv \left( \overline{\epsilon}_1 \gamma^{\mu [\, 4n \,]} \epsilon_2 \right) v_{[\, 4n \,]} \ ,$$
$$\Omega^{[\, 4n-1 \,] I} \equiv 2n \left( \overline{\epsilon}_2 \gamma^{\rho\sigma [\, 4n-1 \,]} \epsilon_1 \right) F_{\rho\sigma}{}^I \ , \tag{2.17}$$
$$\Lambda^I \equiv -\xi^\mu A_\mu{}^I \ .$$

In other words, our system provides non-trivial realization of the algebra

$$\{ Q_\alpha, Q_\beta \} = (\gamma^{\mu [\, 4n \,]})_{\alpha\beta} v_{[\, 4n \,]} P_\mu \ , \tag{2.18}$$

as had been indicated in the past [1][9]. This is also one-particle realization of multiple-particle formulation in higher-dimensions [9], when all the null-vectors in the r.h.s. are replaced by momentum of multi-particles. One of the important relations for the closure check is the Fierz identity for arbitrary Weyl spinors in $D = (9+4n, 1+4n)$:

$$\epsilon_1 \bar{\epsilon}_2 - (1 \leftrightarrow 2) = \frac{1}{2^{4n+3}} \left[ \left( \bar{\epsilon}_1 \gamma^\mu \epsilon_2 \right) \gamma_\mu + \frac{1}{5!} \left( \bar{\epsilon}_1 \gamma^{[5]} \epsilon_2 \right) \gamma_{[5]} + \cdots \right.$$
$$\left. + \frac{1}{(4n+1)!} \left( \bar{\epsilon}_1 \gamma^{[4n+1]} \epsilon_2 \right) \gamma_{[4n+1]} + \frac{1}{(4n+5)!2} \left( \bar{\epsilon}_1 \gamma^{[4n+5]} \epsilon_2 \right) \gamma_{[4n+5]} \right] \ . \quad (2.19)$$

Another technical relation we used for the closure on $\lambda$ is such as

$$\gamma^{\mu\nu[4n]} \gamma^{[4k+1]} \gamma_\mu D_\nu \lambda \ \zeta_{[4k+1]} v_{[4n]}$$
$$= \begin{cases} +2^{4k+4} (4n+1)! D_\nu \lambda \zeta^{[4n]}{}_\nu v_{[4n]} & (\text{for } k = n) \\ 0 & (\text{for } 0 \leq k \leq n-1) \end{cases}, \quad (2.20)$$

where $\zeta_{[m]} \equiv (\bar{\epsilon}_1 \gamma_{[m]} \epsilon_2)$, confirmed as follows: First we separate a single $\gamma$- matrix $\gamma^\mu$ from $\gamma^{\mu\nu[4n]}$, and use the constraint (2.12) and field equation (2.14), being left with the structure $\gamma^{\nu[4n]} \gamma^{[4k+1]} D_\nu \lambda$. We next separate $\gamma^\nu$ out of $\gamma^{\nu[4n]}$ to get $\gamma^{[4n]} \gamma^\nu$, using (2.11), and commute $\gamma^\nu$ with $\gamma^{[4k+1]}$, and use the $\lambda$-field equation. We are now left with the structure $\gamma^{[4n]} \gamma^{[4k]}$. Subsequently, we separate one of the $\gamma$-matrix in $\gamma^{[4n]}$ like $\gamma^{[4n-1]} \gamma^\rho$, and commute $\gamma^\rho$ with $\gamma^{[4k]}$ next to it, using (2.12) again. We repeat this procedure until all the $\gamma$-matrix are used out of $\gamma^{[4k+1]}$ under the constraint (2.12). After all, we get a non-vanishing result, only when $k = n$, yielding the desirable result for our closure.

We now come to the derivation of the $A_\mu$-field equation which had some obstruction beyond $D = (11, 3)$ [11]. This field equation is derived from the $\lambda$-field equation (2.14), as:

$$0 = \delta_Q \left[ \gamma^\mu \left( \partial_\mu \lambda^I + f^{IJK} A_\mu{}^J \lambda^K \right) \right]$$
$$= + \frac{1}{4} \gamma^\mu \gamma^{\rho\sigma[4n]} \epsilon D_\mu F_{\rho\sigma}{}^I v_{[4n]} + f^{IJK} (\bar{\epsilon} \gamma_\mu \lambda^J) \gamma^\mu \lambda^K \quad (2.21)$$
$$= + \frac{1}{2} \gamma^{\sigma[4n]} \epsilon D_\mu F^\mu{}_\sigma{}^I v_{[4n]} + \frac{1}{2^{4n+1}} f^{IJK} \sum_{k=0}^n \frac{(n-k+1)}{(4n+1)!} \gamma^{[4k+1]} \epsilon (\bar{\lambda}^K \gamma_{[4k+1]} \lambda^J) \ .$$

The obstruction pointed out in [11] was that the last summation generates many different structures of $\gamma$-matrces which can not be absorbed into the first $DF$-term with $\gamma^{[4n+1]}$, interpreted as the $A_\mu$-field equation. We can now show that *all* the terms for $0 \leq k \leq n-1$ in (2.21) actually vanish under our constraint (2.12), and only the $k = n$ term survives with the same $\gamma^{[4n+1]}$-structure as the $DF$-term. The proof goes by help of projection operators (2.8), as follows. First, using the constraint (2.12), we see that

$$\not{n}^{(i)} \lambda^I = 0 \implies P_\downarrow^{(i)} \lambda^I = 0 \ ,$$
$$\implies \lambda^I = (P_\uparrow^{(i)} + P_\downarrow^{(i)}) \lambda^I = P_\uparrow^{(i)} \lambda^I \quad (i = 1, 2, \cdots, 4n) \ . \quad (2.22)$$

Now let $S_i$ be the sets of indices for the extra dimensions:

$$S_i \equiv \{9+2i, 10+2i\} \quad (i = 1, \cdots, 4n) \ . \tag{2.23}$$

Next recalling the commutator

$$[P^{(i)}_\downarrow, \gamma_\mu] = \slashed{n}^{(i)} n^{(i)}_\mu - \slashed{m}^{(i)} m^{(i)}_\mu = -[P^{(i)}_\uparrow, \gamma_\mu] \ , \tag{2.24}$$

we see that

$$[P^{(i)}_\downarrow, \gamma^{(j)}_\mu] = 0 \quad (i \neq j) \tag{2.25}$$

for a $\gamma$-matrix $\gamma^{(j)}_\mu$ with $\mu \in S_j$. Now consider $(\overline{\lambda}^K \gamma_{\mu_1 \cdots \mu_{4k+1}} \lambda^J)$ for $0 \leq k \leq n-1$. Since the number $4k+1$ of single $\gamma$-matrices in $\gamma_{\mu_1 \cdots \mu_{4k+1}}$ is smaller than $4n$, there always exists $1 \leq \exists j \leq 4n$, such that

$$\mu_i \notin S_j \quad (1 \leq \forall i \leq 4k+1) \ , \tag{2.26}$$

thus

$$[P^{(j)}_\downarrow, \gamma_{\mu_i}] = 0 \quad (1 \leq \forall i \leq 4k+1) \ . \tag{2.27}$$

Once this is established, we can show for the same $j$ in (2.27) that

$$(\overline{\lambda}^K \gamma_{\mu_1 \cdots \mu_{4k+1}} \lambda^J) = (\overline{\lambda}^K P^{(j)}_\downarrow \gamma_{\mu_1 \cdots \mu_{4k+1}} P^{(j)}_\uparrow \lambda^J) = (\overline{\lambda}^K [P^{(j)}_\downarrow, \gamma_{\mu_1 \cdots \mu_{4k+1}}] P^{(j)}_\uparrow \lambda^J)$$

$$= \left(\overline{\lambda}^K \sum_{l=0}^{4k} \gamma_{[\mu_1 \cdots \mu_l} [P^{(j)}_\downarrow, \gamma_{\mu_{l+1}}] \gamma_{\mu_{l+2} \cdots \mu_{4k+1}]} P^{(j)}_\uparrow \lambda^J\right) = 0 \tag{2.28}$$

holds for $0 \leq \forall k \leq n-1$. This implies that the only non-zero term in (2.21) is for $k = n$:

$$f^{IJK} \gamma^\mu \lambda^K (\overline{\epsilon} \gamma_\mu \lambda^J) = \frac{1}{2^{4n+1}(4n+1)!} f^{IJK} \gamma^{[4n+1]} \epsilon (\overline{\lambda}^K \gamma_{[4n+1]} \lambda^J) \ , \tag{2.29}$$

which can be combined with the $F$-term in (2.21) sharing the common $\gamma^{[4n+1]}$-matrix, yielding our $A_\mu$-field equation (2.15) with one term bilinear in $\lambda$. Hence the obstruction in [11] is circumvented, due to the reduction of the freedom of the $\lambda$-field by our constraint (2.11), resulting in only one source term for the $A_\mu$-field equation.

A proof similar to the one given above holds in $D = (10+4n, 2+4n)$, $(11+4n, 3+4n)$, and $(12+4n, 4+4n)$ as well, so it will not be repeated in the following sections.

Once the component formulation is established, the corresponding superspace formulation in $D = (9+4n, 1+4n)$ is straightforward. We need an additional auxiliary superfield $\chi_{\dot\alpha}$ in addition to $A_\mu$ and $\lambda_\alpha$. Only in superspace formulations, we use the indices $A = (a, \alpha, \dot\alpha)$, $B = (b, \beta, \dot\beta)$, $\cdots$, where $a, b, \cdots = 0, 1, \cdots, 9, 11, 12, \cdots, 10+8n$ for bosonic coordinates, and $\alpha, \beta, \cdots = 1, 2, \cdots, 2^{4n+4}$ or $\dot\alpha, \dot\beta, \cdots = \dot 1, \dot 2, \cdots, \dot 2^{4n+4}$ for fermionic coordinates. This notation is essentially the same as in [1]. The dottedness of $\gamma$-matrices is summarized as

$$C_{\alpha\dot\beta} \ , \quad (\gamma^c)_{\alpha\beta} \ , \quad (\gamma^{[2]})_{\alpha\dot\beta} \ , \quad \cdots \ , \quad (\gamma^{[9+8n]})_{\alpha\beta} \ , \quad (\gamma^{[10+8n]})_{\alpha\dot\beta} \ . \tag{2.30}$$

Our result for superspace constraints is summarized as

$$T_{\alpha\beta}{}^c = (\gamma^{c[4n]})_{\alpha\beta} v_{[4n]} ,\tag{2.31a}$$

$$F_{\alpha b}{}^I = (\gamma_b)_{\alpha\gamma}\lambda^{\gamma I} + (\gamma^{[4n-1]})_\alpha{}^{\dot\beta} \chi_{\dot\beta}{}^I v_{[4n-1]b} ,\tag{2.31b}$$

$$\nabla_\alpha \lambda^{\beta I} = \tfrac{1}{4}(\gamma^{ab[4n]})_\alpha{}^\beta F_{ab}{}^I v_{[4n]} ,\tag{2.31c}$$

$$\nabla_\alpha \chi_{\dot\beta}{}^I = -n(\gamma^{cd})_{\alpha\dot\beta} F_{cd}{}^I ,\tag{2.31d}$$

$$\nabla_\alpha F_{bc}{}^I = (\gamma_{[b}\nabla_{c]}\lambda^I)_\alpha + (\gamma^{[4n-1]}\nabla_{[b|}\chi^I)_\alpha v_{[4n-1]|c]} .\tag{2.31e}$$

Here the multiplication of spinors with $\gamma$-matrices are like $(\gamma^c \chi)_\alpha \equiv (\gamma^c \chi)_\alpha{}^{\dot\beta}\chi_{\dot\beta}$, etc, and the spinorial indices are raised/lowered by $C^{\alpha\dot\beta}$ or $C_{\alpha\dot\beta}$. We do *not* use bars for dotted spinors, as in [1].

The consistency of our system is confirmed by the satisfaction of all the Bianchi identities

$$\nabla_{[A} F_{BC)}{}^I - T_{[AB|}{}^D F_{D|C)}{}^I \equiv 0 ,\tag{2.32}$$

at dimensions up to $d=5/2$, as usual. The important relations we encounter are as follows. At $d=1/2$, we need to show that

$$(\gamma^{d[4n]})_{(\alpha\beta|}(\gamma_d \lambda)_{|\gamma)} v_{[4n]} = 0 ,\tag{2.33}$$

which holds under the constraint (2.11). This can be shown by using the Fierz identity (2.19), and commutators among $\gamma$-matrices. In fact, after the Fierzing, we get

$$(\gamma^{[4k+1]})_{(\alpha\beta|}(\gamma^{[4n]}\gamma_d \gamma_{[4k+1]}\gamma^d \lambda)_{|\gamma)} v_{[4n]}$$
$$= 8(n-k+1)(\gamma^{[4k+1]})_{(\alpha\beta|}(\gamma^{[4n-1]}\gamma^a\gamma_{[4k+1]}\lambda)_{|\gamma)} v_{[4n-1]a} \tag{2.34}$$
$$= 16(n-k+1)(4k+1)(\gamma^{a[4k]})_{(\alpha\beta|}(\gamma^{[4n-1]}\gamma_{[4k]}\lambda)_{|\gamma)} v_{[4n-1]a} ,$$

where we have adopted a method similar to (2.20), namely we separated $\gamma^a$ out of $\gamma^{[4n-1]a}$ and commuted it with $\gamma_{[4k+1]}$, using the constraint (2.12). Repeating the similar procedure, we can show that the l.h.s. of (2.34) is zero, unless $k=n$ or $k=n+1$. However, even the $k=n$ case is shown to equal half the original l.h.s., *i.e.* equal to zero, while the $n=k+1$ case also has zero result due to $(n-k+1)=0$, and therefore (2.34) vanishes for any values of $k=1,2,\cdots,n+1$. At $d=3/2$, we see that (2.31e) satisfies the Bianchi identity, which in turn yields the $\lambda$-field equation, when evaluating each side of the identity $\nabla_{(\alpha}(\nabla_{\beta)}\lambda^{\gamma I}) = \{\nabla_\alpha,\nabla_\beta\}\lambda^{\gamma I}$ [1]. This is nothing but the superspace rewriting of our closure check on $\lambda$ in component. The $A_a$-field equation (2.15) is re-obtained in superspace by evaluating

$$(\gamma_{[4n+1]})^{\alpha\gamma}\nabla_\alpha (\slashed\nabla\lambda^I)_\gamma = 0 .\tag{2.35}$$

The auxiliary superfield $\chi$ completely disappears from all the superfield equations, as in [1]. This concludes the confirmation of our SYM in $D = (9+4n, 1+4n)$.

## 3. SYM in $D = (10+4n, 2+4n)$

The gamma-matrix structure in $D = (10+4n, 2+4n)$ with $n = 0, 1, 2, \cdots$ are exactly parallel to the $D = (10, 2)$ case [1], with the symmetry property [12] and metric

$$\text{Symmetric}: \quad \gamma^{[2]}, \quad \gamma^{[3]}, \quad \cdots, \quad \gamma^{[4k+2]}, \quad \gamma^{[4k+3]}, \quad \cdots, \quad \gamma^{[11+8n]},$$
$$\text{Antisymmetric}: \quad C, \quad \gamma^{\mu}, \quad \cdots, \quad \gamma^{[4k]}, \quad \gamma^{[4k+1]}, \quad \cdots, \quad \gamma^{[12+8n]}, \tag{3.1}$$

$$\left(\eta_{\mu\nu}\right) = \text{diag.} \begin{pmatrix} 0 & 1 & \cdots & 9 & 11 & 12 & 13 & 14 & 15 & 16 & \cdots & 11+8n & 12+8n \\ - & + & \cdots & + & + & - & + & - & + & - & \cdots & + & - \end{pmatrix}. \tag{3.2}$$

Lots of features in this dimensions are parallel to the previous section, whose details are appropriately skipped from now on.

Our supersymmetry transformation rule is summarized as

$$\delta_Q A_\mu{}^I = (\overline{\epsilon}\gamma_\mu \lambda^I) \ ,$$
$$\delta_Q \lambda^I = \tfrac{1}{4} \gamma^{\mu\nu[4n+1]} \epsilon F_{\mu\nu}{}^I v_{[4n+1]} \ . \tag{3.3}$$

$$v_{\mu_1 \cdots \mu_{4n+1}} \equiv n^{(1)}_{[\mu_1} n^{(2)}_{\mu_2} \cdots n^{(4n+1)}_{\mu_{4n+1}]} \ , \tag{3.4}$$

with the null-vectors for $i = 1, 2, \cdots, 4n+1$ defined by

$$\left(n_\mu^{(i)}\right) \equiv \begin{pmatrix} 0 & 1 & \cdots & 9 & 11 & 12 & \cdots & 9+2i & 10+2i & \cdots & 11+8n & 12+8n \\ 0, & 0, & \cdots, & 0, & 0, & 0, & \cdots, & +\tfrac{1}{\sqrt{2}}, & +\tfrac{1}{\sqrt{2}}, & \cdots, & 0, & 0 \end{pmatrix}, \tag{3.5}$$

$$\left(m_\mu^{(i)}\right) \equiv \begin{pmatrix} 0 & 1 & \cdots & 9 & 11 & 12 & \cdots & 9+2i & 10+2i & \cdots & 11+8n & 12+8n \\ 0, & 0, & \cdots, & 0, & 0, & 0, & \cdots, & +\tfrac{1}{\sqrt{2}}, & -\tfrac{1}{\sqrt{2}}, & \cdots, & 0, & 0 \end{pmatrix}. \tag{3.6}$$

As is already clear, the main difference from the $D = (9+4n, 1+4n)$ is the replacement of $4n$ by $4n+1$. The projection operators are defined as in (2.8).

As in $D = (10, 2)$ [1], the gauge field undergoes the extra transformation

$$\delta_E A_\mu{}^I = v_{\mu[4n]} \Omega^{[4n]I} \ , \tag{3.7}$$

and our fields are subject to the constraints

$$v_{[4n]}{}^\mu D_\mu \lambda^I = 0 \ , \tag{3.8}$$

$$v_{[4n]}{}^\mu \gamma_\mu \lambda^I = 0 \ , \tag{3.9}$$

$$v_{[4n]}{}^\mu F_{\mu\nu}{}^I = 0 \ . \tag{3.10}$$

Our gauge field equation has universal shift $4n \to 4n+1$ and a sign change, compared with (2.14) and (2.15):

$$\gamma^\mu D_\mu \lambda^I = 0 \ , \tag{3.11}$$

$$D^\mu F_{\mu[\nu_1}{}^I v_{\nu_2 \cdots \nu_{4n+2}]} = -\frac{1}{2^{4n+1}(4n+2)!} f^{IJK} (\overline{\lambda}^J \gamma_{\nu_1 \cdots \nu_{4n+2}} \lambda^K) \ . \tag{3.12}$$

The closure of our supersymmetry (3.3) is confirmed in the same way as in [1]:

$$[\delta_Q(\epsilon_1), \delta_Q(\epsilon_2)] = \delta_P(\xi) + \delta_G(\Lambda) + \delta_E(\Omega) \ , \tag{3.13}$$

$$\xi^\mu \equiv (\overline{\epsilon}_1 \gamma^{\mu[4n+1]} \epsilon_2) v_{[4n+1]} \ ,$$

$$\Omega^{[4n]\,I} \equiv (2n + \tfrac{1}{2}) (\overline{\epsilon}_2 \gamma^{\rho\sigma[4n]} \epsilon_1) F_{\rho\sigma}{}^I \ , \qquad \Lambda^I \equiv -\xi^\mu A_\mu{}^I \ , \tag{3.14}$$

$$\{Q_\alpha, Q_\beta\} = (\gamma^{\mu[4n+1]})_{\alpha\beta} v_{[4n+1]} P_\mu \ , \tag{3.15}$$

where the $n=0$ case corresponds to [1]. The relevant Fierz identity for arbitrary Weyl spinors is

$$\epsilon_1 \overline{\epsilon}_2 - (1\leftrightarrow 2) = -\frac{1}{2^{4n+4}} \bigg[ \tfrac{1}{2} (\overline{\epsilon}_1 \gamma^{\mu\nu} \epsilon_2) \gamma_{\mu\nu} + \tfrac{1}{6!} (\overline{\epsilon}_1 \gamma^{[6]} \epsilon_2) \gamma_{[6]} + \cdots$$

$$+ \frac{1}{(4n+2)!} (\overline{\epsilon}_1 \gamma^{[4n+2]} \epsilon_2) \gamma_{[4n+2]} + \frac{1}{(4n+6)!2} (\overline{\epsilon}_1 \gamma^{[4n+6]} \epsilon_2) \gamma_{[4n+6]} \bigg] \ . \tag{3.16}$$

Another technical relationship similar to (2.20) is also used to prove the closure on $\lambda$.

In superspace for $D = (9+4n, 1+4n)$ we again need an auxiliary superfield $\chi_\alpha$. The index convention is similar to the previous section, like $A = (a, \alpha, \dot{\alpha})$, $B = (b, \beta, \dot{\beta})$, $\cdots$, where $a, b, \cdots = 0, 1, \cdots, 9, 11, 12, \cdots, 12+8n$ and $\alpha, \beta, \cdots = 1, 2, \cdots, 2^{4n+5}$ or $\dot{\alpha}, \dot{\beta}, \cdots = \dot{1}, \dot{2}, \cdots, \dot{2}^{4n+5}$. The dottedness of $\gamma$-matrices is

$$C_{\alpha\beta} \ , \quad (\gamma^c)_{\alpha\dot{\beta}} \ , \quad (\gamma^{[2]})_{\alpha\beta} \ , \quad \cdots \quad , \quad (\gamma^{[11+8n]})_{\alpha\dot{\beta}} \ , \quad (\gamma^{[12+8n]})_{\alpha\beta} \ . \tag{3.17}$$

In our superspace constraints, we also see the shift of $4n$ to $4n+1$:

$$T_{\alpha\beta}{}^c = (\gamma^{c[4n+1]})_{\alpha\beta} v_{[4n+1]} \ ,$$

$$F_{ab}{}^I = (\gamma_b)_{\alpha\dot{\gamma}} \lambda^{\dot{\gamma} I} + (\gamma^{[4n]})_\alpha{}^\beta \chi_\beta{}^I v_{[4n]b} \ ,$$

$$\nabla_\alpha \lambda^{\dot{\beta} I} = \tfrac{1}{4} (\gamma^{ab[4n+1]})_\alpha{}^{\dot{\beta}} F_{ab}{}^I v_{[4n+1]} \ ,$$

$$\nabla_\alpha \chi_\beta{}^I = (n + \tfrac{1}{4}) (\gamma^{cd})_{\alpha\beta} F_{cd}{}^I \ ,$$

$$\nabla_\alpha F_{bc}{}^I = (\gamma_{[b} \nabla_{c]} \lambda^I)_\alpha + (\gamma^{[4n]} \nabla_{[b|} \chi^I)_\alpha v_{[4n]|c]} \ . \tag{3.18}$$

We can confirm all the Bianchi identities, using relations essentially the same as in the last section. The $A_a$-field equation (3.12) is re-obtained at $d=2$, by evaluating the combination

$$(\gamma_{[4n+2]})^{\alpha\gamma} \nabla_\alpha (\nabla\!\!\!\!/\,\lambda^I)_\gamma = 0 \ . \tag{3.20}$$

## 4. SYM in $D = (11 + 4n, 3 + 4n)$

The chiral structure in $D = (11 + 4n, 3 + 4n)$ with $n = 0, 1, 2 \cdots$ is similar to $D = (9 + 4n, 1 + 4n)$, except that the symmetry property [12] and our metric are now

$$\text{Symmetric}: \quad C, \quad \gamma^{[3]}, \quad \cdots, \quad \gamma^{[4k+3]}, \quad \gamma^{[4k+4]}, \quad \cdots, \quad \gamma^{[12+8n]},$$
$$\text{Antisymmetric}: \quad \gamma^\mu, \quad \gamma^{[2]}, \quad \cdots, \quad \gamma^{[4k+1]}, \quad \gamma^{[4k+2]}, \quad \cdots, \quad \gamma^{[14+8n]}. \quad (4.1)$$

$$\left(\eta_{\mu\nu}\right) = \text{diag.} \left(\begin{smallmatrix}0 & 1 & \cdots & 9 & 11 & 12 & 13 & 14 & 15 & 16 & \cdots & 13+8n & 14+8n \\ - & + & \cdots & + & + & - & + & - & + & - & \cdots & + & -\end{smallmatrix}\right). \quad (4.2)$$

Most of the component equations corresponding to (3.3) - (3.17) are just parallel and self-explanatory. Moreover these are just the generalizations of the $n=0$ case [11], so we list them up with no additional comment, starting with the supersymmetry transformation:

$$\delta_Q A_\mu{}^I = (\bar{\epsilon}\gamma_\mu \lambda^I) ,$$
$$\delta_Q \lambda^I = -\tfrac{1}{4} \gamma^{\mu\nu[4n+2]} \epsilon F_{\mu\nu}{}^I v_{[4n+2]} , \quad (4.3)$$
$$v_{\mu_1 \cdots \mu_{4n+2}} \equiv n^{(1)}_{[\mu_1} n^{(2)}_{\mu_2} \cdots n^{(4n+2)}_{\mu_{4n+2}]} . \quad (4.4)$$

Null Vectors for $i = 1, 2, \cdots, 4n+2$:

$$\left(n^{(i)}_\mu\right) \equiv \left(\begin{smallmatrix}0 & 1 & \cdots & 9 & 11 & 12 & \cdots & 9+2i & 10+2i & \cdots & 13+8n & 14+8n \\ 0, & 0, & \cdots, & 0, & 0, & 0, & \cdots, & +\tfrac{1}{\sqrt{2}}, & +\tfrac{1}{\sqrt{2}}, & \cdots, & 0, & 0\end{smallmatrix}\right), \quad (4.5)$$

$$\left(m^{(i)}_\mu\right) \equiv \left(\begin{smallmatrix}0 & 1 & \cdots & 9 & 11 & 12 & \cdots & 9+2i & 10+2i & \cdots & 13+8n & 14+8n \\ 0, & 0, & \cdots, & 0, & 0, & 0, & \cdots, & +\tfrac{1}{\sqrt{2}}, & -\tfrac{1}{\sqrt{2}}, & \cdots, & 0, & 0\end{smallmatrix}\right). \quad (4.6)$$

Extra Transformation:

$$\delta_E A_\mu{}^I = v_{\mu[4n+1]} \Omega^{[4n+1]I} . \quad (4.7)$$

Constraints:

$$v_{[4n+1]}{}^\mu D_\mu \lambda^I = 0 , \quad (4.8)$$
$$v_{[4n+1]}{}^\mu \gamma_\mu \lambda^I = 0 , \quad (4.9)$$
$$v_{[4n+1]}{}^\mu F_{\mu\nu}{}^I = 0 . \quad (4.10)$$

Field Equations:

$$\gamma^\mu D_\mu \lambda^I = 0 , \quad (4.11)$$
$$D^\mu F_{\mu[\nu_1}{}^I v_{\nu_2 \cdots \nu_{4n+3}]} = \frac{1}{2^{4n+2}(4n+3)!} f^{IJK} (\bar{\lambda}^J \gamma_{\nu_1 \cdots \nu_{4n+3}} \lambda^K) . \quad (4.12)$$

Closure of Gauge Algebra:

$$[\delta_Q(\epsilon_1), \delta_Q(\epsilon_2)] = \delta_P(\xi) + \delta_G(\Lambda) + \delta_E(\Omega) \ , \tag{4.13}$$

$$\xi^\mu \equiv -\left(\bar{\epsilon}_1 \gamma^{\mu[4n+2]} \epsilon_2\right) v_{[4n+2]} \ ,$$

$$\Omega^{[4n+1]I} \equiv -(2n+1)\left(\bar{\epsilon}_2 \gamma^{\rho\sigma[4n+1]} \epsilon_1\right) F_{\rho\sigma}{}^I \ , \qquad \Lambda^I \equiv -\xi^\mu A_\mu{}^I \ , \tag{4.14}$$

$$\{Q_\alpha, Q_\beta\} = (\gamma^{\mu[4n+2]})_{\alpha\beta} v_{[4n+2]} P_\mu \ . \tag{4.15}$$

Fierz Identity for Weyl Spinors:

$$\epsilon_1 \bar{\epsilon}_2 - (1 \leftrightarrow 2) = -\frac{1}{2^{4n+5}} \left[ \frac{1}{3!} \left(\bar{\epsilon}_1 \gamma^{[3]} \epsilon_2\right) \gamma_{[3]} + \frac{1}{7!} \left(\bar{\epsilon}_1 \gamma^{[7]} \epsilon_2\right) \gamma_{[7]} + \cdots \right.$$
$$\left. + \frac{1}{(4n+3)!} \left(\bar{\epsilon}_1 \gamma^{[4n+3]} \epsilon_2\right) \gamma_{[4n+3]} + \frac{1}{(4n+6)!2} \left(\bar{\epsilon}_1 \gamma^{[4n+7]} \epsilon_2\right) \gamma_{[4n+7]} \right] \ . \tag{4.16}$$

An additional auxiliary superfield $\chi_\alpha$ is needed in superspace, where index convention is $A = (a, \alpha, \dot{\alpha})$, $B = (b, \beta, \dot{\beta})$, $\cdots$, with $a$, $b$, $\cdots = 0, 1, \cdots, 9, 11, 12, \cdots, 14+8n$ and $\alpha, \beta, \cdots = 1, 2, \cdots, 2^{4n+6}$ or $\dot{\alpha}, \dot{\beta}, \cdots = \dot{1}, \dot{2}, \cdots, \dot{2}^{4n+6}$. The results in superspace are summarized as:

Dottedness:

$$C_{\alpha\dot{\beta}} \ , \quad (\gamma^c)_{\alpha\beta} \ , \quad (\gamma^{[2]})_{\alpha\dot{\beta}} \ , \quad \cdots \ , \quad (\gamma^{[13+8n]})_{\alpha\beta} \ , \quad (\gamma^{[14+8n]})_{\alpha\dot{\beta}} \ . \tag{4.17}$$

Superspace Constraints:

$$T_{\alpha\beta}{}^c = (\gamma^{c[4n+2]})_{\alpha\beta} v_{[4n+2]} \ ,$$
$$F_{\alpha b}{}^I = -(\gamma_b)_{\alpha\gamma} \lambda^{\gamma I} + (\gamma^{[4n+1]})_\alpha{}^\beta \chi_\beta{}^I v_{[4n+1]b} \ ,$$
$$\nabla_\alpha \lambda^{\beta I} = \tfrac{1}{4} \left(\gamma^{ab[4n+2]}\right)_\alpha{}^\beta F_{ab}{}^I v_{[4n+2]} \ ,$$
$$\nabla_\alpha \chi_\beta{}^I = -\left(n + \tfrac{1}{2}\right)(\gamma^{cd})_{\alpha\beta} F_{cd}{}^I \ . \tag{4.18}$$

Superfield Equation for $A_a$:

$$(\gamma_{[4n+3]})^{\alpha\gamma} \nabla_\alpha \left(\nabla\!\!\!\!/ \lambda^I\right)_\gamma = 0 \ . \tag{4.19}$$

## 5. SYM in $D = (12 + 4n, 4 + 4n)$

The $\gamma$-matrix structure in $D = (12 + 4n, 4 + 4n)$ with $n = 0, 1, 2, \cdots$ is similar to that in $D = (10 + 4n, 2 + 4n)$. Our component results are:

Symmetry:

$$\text{Symmetric}: \ C \ , \ \gamma^\mu \ , \ \cdots \ , \ \gamma^{[4k]} \ , \ \gamma^{[4k+1]} \ , \ \cdots \ , \ \gamma^{[16+8n]} \ ,$$
$$\text{Antisymmetric}: \ \gamma^{[2]} \ , \ \gamma^{[3]} \ , \ \cdots \ , \ \gamma^{[4k+2]} \ , \ \gamma^{[4k+3]} \ . \ \cdots \ , \ \gamma^{[15+8n]} \ , \tag{5.1}$$

Metric:

$$\left(\eta_{\mu\nu}\right) = \mathrm{diag.} \left( \overset{0}{-} \ \overset{1}{+} \ \overset{\cdots}{\cdots} \ \overset{9}{+} \ \overset{11}{+} \ \overset{12}{+} \ \overset{13}{-} \ \overset{14}{+} \ \overset{15}{-} \ \overset{16}{+} \ \overset{\cdots}{-} \ \overset{15+8n}{\cdots} \ \overset{16+8n}{+} \ - \right). \quad (5.2)$$

Supersymmetry Transformation Rule:

$$\delta_Q A_\mu{}^I = (\bar{\epsilon}\gamma_\mu \lambda^I) ,$$
$$\delta_Q \lambda^I = -\tfrac{1}{4} \gamma^{\mu\nu[4n+3]} \epsilon F_{\mu\nu}{}^I v_{[4n+3]} , \quad (5.3)$$
$$v_{\mu_1 \cdots \mu_{4n+3}} \equiv n^{(1)}_{[\mu_1} n^{(2)}_{\mu_2} \cdots n^{(4n+3)}_{\mu_{4n+3}]} . \quad (5.4)$$

Null Vectors for $i = 1, 2, \cdots, 4n+3$:

$$\left(n^{(i)}_\mu\right) \equiv \left( \overset{0}{0}, \ \overset{1}{0}, \ \overset{\cdots}{\cdots}, \ \overset{9}{0}, \ \overset{11}{0}, \ \overset{12}{0}, \ \overset{\cdots}{\cdots}, \ \overset{9+2i}{+\tfrac{1}{\sqrt{2}}}, \ \overset{10+2i}{+\tfrac{1}{\sqrt{2}}}, \ \overset{\cdots}{\cdots}, \ \overset{15+8n}{0}, \ \overset{16+8n}{0} \right), \quad (5.5)$$

$$\left(m^{(i)}_\mu\right) \equiv \left( \overset{0}{0}, \ \overset{1}{0}, \ \overset{\cdots}{\cdots}, \ \overset{9}{0}, \ \overset{11}{0}, \ \overset{12}{0}, \ \overset{\cdots}{\cdots}, \ \overset{9+2i}{+\tfrac{1}{\sqrt{2}}}, \ \overset{10+2i}{-\tfrac{1}{\sqrt{2}}}, \ \overset{\cdots}{\cdots}, \ \overset{15+8n}{0}, \ \overset{16+8n}{0} \right). \quad (5.6)$$

Extra Transformation:

$$\delta_E A_\mu{}^I = v_{\mu[4n+2]} \Omega^{[4n+2]I} . \quad (5.7)$$

Constraints:

$$v_{[4n+2]}{}^\mu D_\mu \lambda^I = 0 , \quad (5.8)$$
$$v_{[4n+2]}{}^\mu \gamma_\mu \lambda^I = 0 , \quad (5.9)$$
$$v_{[4n+2]}{}^\mu F_{\mu\nu}{}^I = 0 . \quad (5.10)$$

Field Equations:

$$\gamma^\mu D_\mu \lambda^I = 0 , \quad (5.11)$$
$$D^\mu F_{\mu[\nu_1}{}^I v_{\nu_2 \cdots \nu_{4n+4}]} = \frac{1}{2^{4n+3}(4n+4)!} f^{IJK} (\bar{\lambda}^J \gamma_{\nu_1 \cdots \nu_{4n+4}} \lambda^K) . \quad (5.12)$$

Closure of Gauge Algebra:

$$[\delta_Q(\epsilon_1), \delta_Q(\epsilon_2)] = \delta_P(\xi) + \delta_G(\Lambda) + \delta_E(\Omega) , \quad (5.13)$$
$$\xi^\mu \equiv - \left(\bar{\epsilon}_1 \gamma^{\mu[4n+3]} \epsilon_2\right) v_{[4n+3]} ,$$
$$\Omega^{[4n+2]I} \equiv - \left(2n + \tfrac{3}{2}\right) \left(\bar{\epsilon}_2 \gamma^{\rho\sigma[4n+2]} \epsilon_1\right) F_{\rho\sigma}{}^I , \quad \Lambda^I \equiv -\xi^\mu A_\mu{}^I , \quad (5.14)$$
$$\{Q_\alpha, Q_\beta\} = (\gamma^{\mu[4n+3]})_{\alpha\beta} v_{[4n+3]} P_\mu . \quad (5.15)$$

Fierz Identity for Weyl Spinors:

$$\epsilon_1 \bar{\epsilon}_2 - (1 \leftrightarrow 2) = \frac{1}{2^{4n+6}} \left[ (\bar{\epsilon}_1 \epsilon_2) + \tfrac{1}{4!} \left(\bar{\epsilon}_1 \gamma^{[4]} \epsilon_2\right) \gamma_{[4]} + \cdots \right.$$
$$\left. + \tfrac{1}{(4n+4)!} \left(\bar{\epsilon}_1 \gamma^{[4n+4]} \epsilon_2\right) \gamma_{[4n+8]} + \tfrac{1}{(4n+8)!2} \left(\bar{\epsilon}_1 \gamma^{[4n+8]} \epsilon_2\right) \gamma_{[4n+8]} \right] . \quad (5.16)$$

The superspace formulation with the indices $A = (a, \alpha, \dot\alpha)$, $B = (b, \beta, \dot\beta)$, $\cdots$, $a, b, \cdots = 0, 1, \cdots, 9, 11, 12, \cdots, 16+8n$ and $\alpha, \beta, \cdots = 1, 2, \cdots, 2^{4n+7}$ or $\dot\alpha, \dot\beta, \cdots = \dot 1, \dot 2, \cdots, \dot 2^{4n+7}$ is summarized as:

Dottedness:

$$C_{\alpha\beta}, \quad (\gamma^c)_{\alpha\dot\beta}, \quad (\gamma^{[2]})_{\alpha\beta}, \quad \cdots, \quad (\gamma^{[15+8n]})_{\alpha\dot\beta}, \quad (\gamma^{[16+8n]})_{\alpha\beta}. \tag{5.17}$$

Superspace Constraints:

$$\begin{aligned}
T_{\alpha\beta}{}^c &= \left(\gamma^{c[4n+3]}\right)_{\alpha\beta} v_{[4n+3]}, \\
F_{ab}{}^I &= -(\gamma_b)_{\alpha\dot\gamma} \lambda^{\dot\gamma I} + \left(\gamma^{[4n+2]}\right)_\alpha{}^\beta \chi_\beta v_{[4n+2]b}, \\
\nabla_\alpha \lambda^{\dot\beta I} &= \tfrac{1}{4} \left(\gamma^{ab[4n+3]}\right)_\alpha{}^{\dot\beta} F_{ab}{}^I v_{[4n+3]}, \\
\nabla_\alpha \chi_\beta{}^I &= \left(n + \tfrac{3}{4}\right)(\gamma^{cd})_{\alpha\beta} F_{cd}{}^I.
\end{aligned} \tag{5.18}$$

Superfield Equation for $A_a$:

$$(\gamma_{[4n+4]})^{\alpha\gamma} \nabla_\alpha \left(\not\!\nabla \lambda^I\right)_\gamma = 0. \tag{5.19}$$

## 6. Concluding Remarks

In this paper we have constructed non-trivial interacting SYM theories in arbitrary higher even dimensions $D = (9+m, 1+m)$ where $m = 0, 1, 2, \cdots$. We have categorized the whole set of even dimensional space-time of this sort into four classes, depending on the integer $m$ modulo 4, based on the $\gamma$-matrix structures [12]. We see that there is no obstruction for each of these classes, due to the general property of the null-vectors and projection operators, as well as the common features of gamma-matrix algebra, combined with our constraints on fields. Even though the usage of null-vectors violates the manifest Lorentz symmetry in these higher-dimensions, this feature is exactly the same as in the $D = (10, 2)$ [1] and $D = (11, 3)$ [11] cases.

The existence of SYM theory in any arbitrary higher even dimensions suggests possible supersymmetry formulation with *infinitely* many superparticles, each carrying different proper time coordinates, like supermultiple-time formulation [13] in the second quantization of field theory. In fact, we can define the algebra with arbitrary number of massless particles

$$\{Q_\alpha, Q_\beta\} = (\gamma^{\mu\nu_1 \cdots \nu_m})_{\alpha\beta} P_{0\mu} P_{1\nu_1} \cdots P_{m\nu_m}, \tag{6.1}$$

as generalization of refs. [9][11] for $m = 0, 1, \cdots$, and subsequently take the limit $m \to \infty$. It should be also straightforward to generalize the results in [8][14] for superparticles in arbitrarily higher dimensions. In fact, a recent superparticle formulation in [15] seems easily generalized in higher dimensions even for massive superparticles. It is interesting to notice that the $\gamma$-matrix structure changes every two dimensions modulo eight dimensions or equivalently every four particles, due to the property of Clifford algebra [12].

We have shown that there is no limit for dimensionalities for SYM theories, once null-vectors are introduced. This is against the common expectation for the maximal dimensions even for Lorentz non-invariant formulation, as $D = (9,1)$ was the maximum for Lorentz invariant case. Even though we dealt only with higher even dimensions here, we believe that the formulations in odd dimensions also work, as expected from dimensional reductions. The existence of consistent SYM theories in arbitrary higher even dimensions suggests certain underlying huge class of dualities that had never been known before in superstring/p-brane physics. Our result indicates also the existence of consistent supergravity theories in any arbitrary higher dimensions beyond $D = (10,1)$, once null-vectors are introduced into the formulation. Studies in these directions are now under way [16]

We are grateful to I. Bars, S.J. Gates, Jr. and C. Vafa for important discussions. Special acknowledgement is for E. Sezgin who gave lots of suggestions to improve the manuscripts.